\documentclass[a4paper,10pt]{article}

\date{}

\usepackage[dvips]{graphicx}
\usepackage{amsmath}
\usepackage{citesort}
\usepackage{floatflt}  

\reversemarginpar
\usepackage{bm}

\title{Isobar configurations: $\Delta N$ correlations\\ versus the independent particle model}
\author{I. V. Glavanakov, A. N. Tabachenko \\
\it Institute of Physics and Technology, Tomsk Polytechnic University, \\
\it Tomsk, Russia}

\begin{document}

\maketitle

\begin{center}
\noindent
\parbox{15cm}
{\small
We present a comparative analysis of two models for the $A( {\gamma ,\,\pi
N})B$ reaction, which take into account the isobar configurations in
the ground state of the nuclei: the $\Delta N$ correlation model and the
 quasifree pion photoproduction model. The considered models differ in their
 descriptions of the nucleus states. The $\Delta N$ correlation
model takes into account the dynamic correlations of the nucleon and isobar
formed in the virtual transition $NN \to \Delta N$, and in the
quasifree pion photoproduction model, isobars and nucleons in the nucleus are
considered as independent constituents. The predictions of the models are
considered for two reactions ${}^{16}$O$( {\gamma ,\,\pi ^{ +} p}
){}^{15}$C and ${}^{16}$O$( {\gamma ,\,\pi ^{ -} p} ){}^{15}$O.
It is shown, that the two models predict the differential cross section
significantly differing both in  absolute values, and in the shape of the angular
dependence. We compare the results of the $\Delta N$ correlation model for
the $( {\gamma ,\,\pi N} )$ and $( {\gamma ,\,\pi NN})$
reactions with the ${}^{16}$O$( {\gamma ,\, \pi ^{ -} p})$ reaction data
measured at BNL. Our results give support to the $\Delta N$ correlation
model.
}\\
\end{center}

\section{Introduction}
For a long time, the role played by nucleon resonances as components of the
atomic nucleus was intensively studied both experimentally and theoretically
\cite{1,2}. According to theoretical estimates, some  of the nucleons in the
nucleus, as a result of collisions, can experience excitation of the internal
degrees of freedom and go with a probability of a few percent to a virtual
isobar states \cite{3,4}. To describe such nuclear states, the wave function of
the nucleus, including the nucleon configurations, is complemented by isobar
configurations, in which one or more nucleons are in an excited states.
Consideration of the isobar configurations in the ground state of the nuclei is
important in explaining both the static properties of the nuclei and the
nuclear reactions.

Nuclear reactions that cannot be explained within a model that assumes a
single interaction of a projectile particle with bound nucleons of a nucleus
are an efficient tool in experimentally studying isobar degrees of freedom
in the ground states of nuclei. As an example, we can indicate $( {\pi ^{ + },\,\pi ^{ -} p})$
reactions \cite{5,6}, where the charged state of a
scattered particle changes by $2e$, or $( {p,\, p'\,\pi ^{ +} p})$
\cite{7} and $( {\gamma ,\,\pi ^{ -} n})$ reactions accompanied by the
production of particles whose total electric charge is +2 or $-$1. Such
experimental data are usually interpreted, using the model of the quasifree
knockout of the isobar \cite{5,6,7,8,9}. The weakest element of this approach is the
independent particle model, used as a model of the nucleus. Because the
virtual isobar is formed in the nucleus owing to the $NN \to \Delta N$ and
$NN \to \Delta \Delta $ transitions, the states of the nucleon and isobar of
the $\Delta N$ system or the states of two isobars of the $\Delta \Delta $
systems are interdependent. The independent particle model does not account for
these dynamic correlations that may cause distortion of the theoretical
predictions and inadequate interpretation of the experimental data.

Recently, we proposed a model of the $A( {\gamma ,\,\pi N})B$
reaction that takes into account the $\Delta N$ correlations of the nuclear
wave function \cite{10}. The $\Delta N$ correlation model sequentially considers
production of the virtual $\Delta$-isobar in the nucleus and its participation in the
production of the pion-nucleon pair. The model includes both direct and
exchange reaction mechanisms. In this paper, we present the comparative
analysis of the $\Delta N$ correlation model for the $A( {\gamma ,\,\pi N})B$ 
reaction and the quasifree pion photoproduction model.

Currently, there are no exclusive experimental data for the $A( {\gamma ,\,\pi
N})B$ reaction, measured at the high momenta of the residual nucleus,
where the contribution of the isobar configurations in the reaction cross section
can be expected to be significant. Available data include the contribution
of the final states in which the residual nucleus is disintegrated. We used
the $\Delta N$ correlation approach for the analysis of such data  \cite{11}.
In the same way as the short-range nucleon-nucleon correlations were the
starting point to explain the $( {e,\, e'NN})$ reactions, the
$\Delta N$ correlation served in \cite{12} as a basis for the model of the $(
{\gamma ,\pi NN})$ reaction --  pion photoproduction with the emission
of two nucleons. Using the $\Delta N$ correlation model of the $( {\gamma
,\,\pi N})$ and $( {\gamma ,\,\pi NN})$ reactions, the
${}^{16}$O$( {\gamma ,\, \pi ^{ +} p})$ reaction data were
interpreted in \cite{11}. 
In the present paper, this approach is used by us to analyze
the ${}^{16}$O$( {\gamma ,\, \pi ^{ -} p})$ reaction data
measured at the Laser Electron Gamma Source (LEGS) facility of Brookhaven National
Laboratory (BNL) \cite{13}.

\section{Models for pion photoproduction from nuclei}

In the framework of the formalism developed in \cite{10} the squared modulus of
the direct amplitude \textit{T}$_{d}$ of the reaction $A( {\gamma
,\, \pi N})B$, summed over the states \textit{f}$_{B}$ of the
residual nucleus \textit{B}, can be written as
\begin{equation}
\label{eq1}
\sum\limits_{f_{B}}  {|
{T_{d}}|^{2}} = A\int {d( {X_{1} ',X_{1} ,\tilde {X}_{1}
',\tilde {X}_{1}})\,}  \Phi _{\alpha} ^{\ast} ( {X_{1} '}
) < X_{1} '|t_{\gamma \pi}  |X_{1} > \rho( {X_{1} ;\tilde
{X}_{1}}) < \tilde {X}_{1} |t_{\gamma \pi} ^{ \dagger}  |\tilde {X}_{1} '
> \Phi _{\alpha}( {\tilde {X}_{1} '}),
\end{equation}
where $t_{\gamma \pi}  $ is the single-particle operator of the pion
photoproduction on free baryons and $\Phi _{\alpha}   $ is the wave
function of the free nucleon in the state $\alpha$,
\begin{equation}
\rho ( {X_{1} ;\tilde {X}_{1}}) = \int {d( {X_{2}
,...,X_{A}}) }  \Psi _{\beta} ( {X_{1} ,X_{2}
,...,X_{A}})
 \Psi _{\beta} ^{\ast}( {\tilde {X}_{1} ,X_{2}
 ,...,X_{A}}) \label{eq2}
\end{equation}
is the one-body density matrix and $\Psi _{\beta}   $ is the wave function of
the nucleus. Here we use the approach developed in \cite{3}, according to which 
baryon bound in the nucleus, in addition to the space \textbf{r}, spin
\textit{s}, and isospin \textit{t} coordinates ($\textbf{r},s,t \equiv x$), is
characterized also by the intrinsic coordinate \textit{m} $( {x,m
\equiv X})$, which specifies the position of a baryon in a space of  
intrinsic states. In (\ref{eq1}) and (\ref{eq2}), the integral sign denotes the integration
over continuous variables and summation over discrete variables.

As can be seen from (\ref{eq1}), the operator of the pion photoproduction $t_{\gamma
\pi}  $ and the wave function $\Psi _{\beta}   $ of the \textit{A}
nucleus, defining the one-body density matrix $\rho ( {X_{1} ;\tilde
{X}_{1}})$, are two main components of the reaction model.

Nuclear wave function in general is the superposition of different
configurations and may be represented as
\[
\Psi _{\beta}  ( {X_{1} ,...,X_{A}}) = \sum\limits_{n} {A_{n}
}  \varphi _{n} ( {m_{1} ,...,m_{A}} )  \psi _{\beta} ^{n}
( {x_{1} ,...,x_{A}} ),
\]
\noindent
where $\psi _{\beta} ^{n} ( {x_{1} ,...,x_{A}} )$ is the wave
function describing the state of \textit{A} baryons in the usual, spin and
isotopic spaces, $\varphi _{n} ( {m_{1} ,...,m_{{\rm A}}})$ is
the wave function describing the intrinsic state of the baryons \cite{3}. The
index $\beta \equiv \beta _{1} ,...,\beta _{A} $ characterizes the usual
space, spin and isospin states of \textit{A} particles. The index $n \equiv
n_{1} ,...,n_{A} $ defines the intrinsic states of the particles. The
particle can be a nucleon (\textit{n}$_{i}$\textit{ = N}), $\Delta$-isobar
(\textit{n}$_{i}$ = $\Delta$) or other excited states of the nucleon. $A_{n} $ is
the antisymmetrization operator. The free nucleon wave function in such
an approach can be written as $\Phi _{\alpha}  ( {X}) = \varphi
_{N} ( {m} )  \phi _{\alpha _{n}}  ( {x})$, where
$\alpha _{n} \equiv \textbf{p}_{n} ,m\sigma _{n} ,m\tau _{n} $ is the index of the
nucleon state with momentum \textbf{p}$_{n}$, spin projection on the
selected direction $m\sigma _{n} $ and the third isospin projection $m\tau
_{n} $.

The wave function $\Psi _{\beta}   $ satisfies the Schr\"{o}dinger
equation
\begin{equation}
\label{eq3}
( {H - E_{\beta} } ) \Psi _{\beta} = 0,
\end{equation}
\noindent
where the Hamiltonian of the baryon system $H$ may be represented 
as $H = H_{0} +   V$.
The operator $H_{0} $ includes the kinetic energy operator 
and part associated with the internal degrees of freedom 
\cite{3} and $V = \sum\nolimits_{i < j} {V_{ij}}  $, $V_{ij} $ is the interaction potential of   
the \textit{i}-th and \textit{j}-th particle.

In the considered models of the $A( {\gamma ,\, \pi N})B$
reaction, the nuclear wave function $\Psi _{\beta}   $ includes two
intrinsic configurations
\[
\Psi _{\beta}   = \Psi _{\beta} ^{N} + \Psi _{\beta} ^{\Delta}  :
\]
\noindent
a configuration, in which all particles are nucleons 
\[
\Psi _{\beta} ^{N} ( {X_{1} ,...,X_{A}})  =  \varphi
_{n_{N}}  ( {m_{1} ,...,m_{{\rm A}}})\psi _{\beta} ^{n_{N}}
( {x_{1} ,...,x_{A}}),
\]
\noindent
where the index $n_{N} \equiv N,N,...,N$, and an isobar configuration
\[
\Psi _{\beta} ^{\Delta} ( {X_{1} ,...,X_{A}})  =
 A_{n_{\Delta} }   \varphi _{n_{\Delta} }  ( {m_{1} ,...,m_{{\rm
A}}})\psi _{\beta} ^{n_{\Delta} } ( {x_{1} ,...,x_{A}}
),
\]
\noindent
in which one particle is an $\Delta$-isobar, and the rest are nucleons. Here the
index $n_{\Delta}  \equiv \Delta ,N,...,N$.

In the following, we give the comparative analysis of the $\Delta
N$ correlation model for the direct reaction mechanisms of the $A(
{\gamma ,\,\pi N} )B$ reaction and the quasifree pion
photoproduction model. We will start with a detailed examination of the $\Delta
N$ correlation model. Assuming that only two nucleons are involved in the
excitation of the nucleon$^{\prime}$s  internal degrees of freedom, the wave function $\Psi
_{\beta} ^{\Delta}  ( {X_{1} ,...,X_{A}}) $ of the isobar
configuration can be written as the superposition of the products of the
wave function $\Psi _{[ {\beta _{i} \beta _{j}}]}^{\Delta N}
( {X_{1} ,X_{2}})$ of the $\Delta$\textit{N} system, which includes an
isobar and the second nucleon (the participant of the transition \textit{NN}$\rightarrow
\Delta$\textit{N}) and the wave function $\Psi _{( {\beta _{i} \beta _{j}
})^{ - 1}}^{N} ( {X_{3} ,...,X_{A}})$, describing the
state of the nucleon core, which includes other \textit{A}--2 nucleons,
\begin{equation}
\Psi _{\beta} ^{\Delta}  ( {X_{1} ,...,X_{A}} )  = 
 A_{\Delta  }   \sum\limits_{ij} {\Psi _{[ {\beta _{i} \beta _{j}
}]}^{\Delta N}( {X_{1} ,X_{2}})} 
 \Psi _{( {\beta
_{i} \beta _{j}})^{ - 1}}^{N}( {X_{3} ,...,X_{A}}) \nonumber.
\end{equation}
\noindent Here
\[
 \Psi _{[ {\beta _{i} \beta _{j}} ]}^{\Delta N}( {X_{1}
,X_{2}})  =   A_{\Delta N}  \,  \varphi _{\Delta N}(
{m_{1} ,m_{2}})  \psi _{[ {\beta _{i} \beta _{j}}
]}^{\Delta N}( {x_{1} ,x_{2}}),
\]
\begin{equation}
 \Psi _{( {\beta _{i} \beta _{j}})^{ - 1}}^{N}( {X_{3}
,...,X_{A}}) = \ 
\varphi _{n_{N}} ( {m_{3} ,...,m_{A}}
) \psi _{( {\beta _{i} \beta _{j}})^{ - 1}}^{n_{N}}
( {x_{3} ,...,x_{A}}), \nonumber
\end{equation}
\noindent $A_{\Delta}   $ and $A_{\Delta N}  $ are the antisymmetrization operators
of the wave functions.

The equation for the wave function of the $\Delta$\textit{N} system $\psi _{[
{\beta _{i} \beta _{j}}]}^{\Delta N} $ was obtained from equation
(\ref{eq3}), using the diagonality of the operator $H_{0} $,
\[
 < n_{\Delta}  |( {H - E_{\beta} } )A_{\Delta}  A_{\Delta N}
|n_{\Delta}  > \sum\limits_{ij} {\psi _{[ {\beta _{i} \beta _{j}}
]}^{\Delta N}( {x_{1} ,x_{2}} ) }  \psi _{( {\beta
_{i} \beta _{j}})^{ - 1}}^{n_{N}}( {x_{3} ,...,x_{A}}
) = - < n_{\Delta}  |V|n_{N} > \psi _{\beta} ^{n_{N}}( {x_{1}
,...,x_{A}}).
\]
\begin{figure*}
\centering
\includegraphics[width=0.85\textwidth,keepaspectratio]{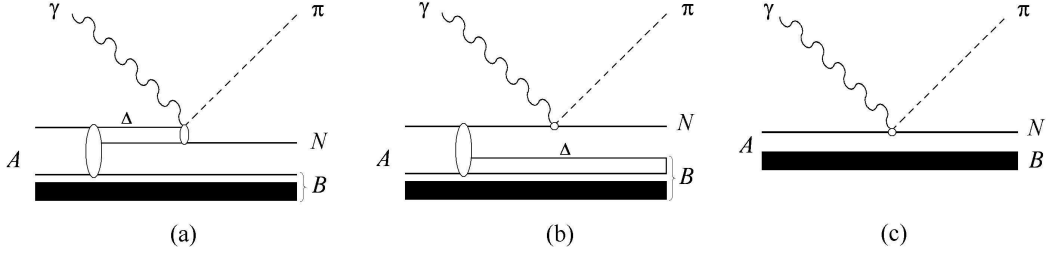}
\vspace*{4mm}\caption{\label{Fig:1}\small Diagrams illustrating the direct mechanisms
of the pion production in the $A( {\gamma ,\, \pi N})B$ reaction.}
\end{figure*}
  The wave functions of the bound nucleon systems $\psi  ^{n_{N}}  $ were
calculated in the framework of the harmonic oscillator shell model, which
reproduces the mean square charge radius of the nucleus.

A one-particle density matrix $\rho ( {X_{1} ;\tilde {X}_{1}})$
was analyzed in \cite{10}, taking into account the isobar configuration of the
nuclear wave function. According to \cite{10} direct mechanisms of the reaction
$A( {\gamma ,\, \pi N})B$ are caused by the following
components of the density matrix
\begin{equation} \label{eq4}
\rho  = \rho ^{\Delta}  + \rho ^{N} + \rho ^{C},	
\end{equation}
\noindent where
\begin{equation} \label{eq5}
\rho  ^{\Delta}  ( {X_{1} ;\tilde {X}_{1}} ) = \varphi
_{\Delta}( {m_{1}} )\left[  {\frac{{1}}{{A}}\sum\limits_{ij}
{\int {dx_{2}}\,  \psi _{[ {\beta _{i} \beta _{j}}]}^{\Delta N}
( {x_{1} ,x_{2}})
\psi _{[ {\beta _{i} \beta _{j}}
]}^{\Delta N\ast} ( {\tilde {x}_{1} ,x_{2}} )}}
\right]\varphi _{\Delta} ^{\ast} ( {\tilde {m}_{1}} ),
\end{equation}
\[
\rho ^{N}( {X_{1} ;\tilde {X}_{1}}) = \varphi _{N} (
{m_{1}} )\left[ {\frac{{1}}{{A}}\sum\limits_{ij}  {\int {dx_{2}} \,
\psi _{[ {\beta _{i} \beta _{j}}]}^{\Delta N}( {x_{2}
,x_{1}}) \psi _{[ {\beta _{i} \beta _{j}}]}^{\Delta
N\ast} ( {x_{2} ,\tilde {x}_{1}})}}  \right]\varphi _{N}^{\ast
} ( {\tilde {m}_{1}} ),
\]
\[
\rho ^{C}( {X_{1} ;\tilde {X}_{1}}) = \varphi _{N}({m_{1}})
\left[ \frac{{1}}{{A}} \left(
N_N \sum\limits_{i=1}^A
\psi _{\beta _{i}} ( x_{1})\, \psi _{\beta _{i}}^* (\tilde x_{1})
+ \sum\limits_{ij,k\neq ij} N_{\Delta ij}\,
\psi _{\beta _{k}}( x_{1} )\, \psi _{\beta _{k}}^* (\tilde x_{1} )
\right) \right]
\varphi _{N}^{\ast}( {\tilde {m}_{1}} ).
\]
\noindent Here, $\psi _{\beta}   ( {x})$ is the single-particle wave
function of the bound nucleon in the nucleus in a state $\beta$ and $N_{\Delta}  =
\sum\nolimits_{ij} {N_{\Delta  ij}}  $,\textit{} $N_{N} $ are the norms of
the wave functions $\Psi _{\beta} ^{\Delta}  $ and $\Psi _{\beta} ^{N} $.

In the context of equation (\ref{eq1}), these components (\ref{eq4}) of the density matrix
correspond to the reaction mechanisms, which are illustrated by the diagrams
in Figs. \ref{Fig:1}\textit{a}, \ref{Fig:1}\textit{b} and \ref{Fig:1}\textit{c}.

The diagrams in Figs. 1\textit{a} and 1\textit{b} describe the mechanisms of
the reactions, in which the production of the pion--nucleon pair occurs at
the interaction of photon with the isobar and nucleon of the $\Delta$\textit{N}
system. The pion production as a result of the mechanism, corresponding to
 diagram 1\textit{c}, occurs at the interaction of a photon with a nucleon of
the nucleon core. In this case, the wave function of the residual nucleus
\textit{B} includes both the nucleon and isobar configurations.%

\begin{figure*}
\centering
\includegraphics[width=0.7\textwidth,keepaspectratio]{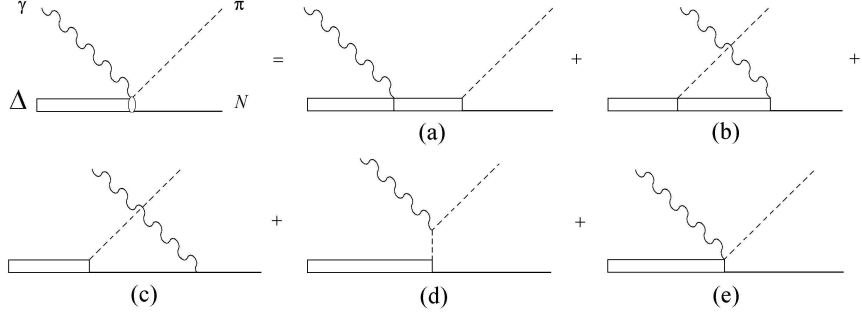}
\vspace*{4mm}\caption{\label{Fig:2}\small  Diagrams for describing the $\gamma \Delta \to N\pi $ process.}
\end{figure*}

Consider the second component of the model -- the operator $t_{\gamma \pi}
$. The operators of the pion production, acting in the spaces of the
coordinates \textit{x} and \textit{X}, are related by the equation
$$
< x'|t_{\gamma B \to N\pi}  |x >   = \sum\limits_{{m}',m} {\varphi
_{N}^{\ast}  ( {{m}'})} < X'|t_{\gamma \pi}   |X > \varphi
_{B} ( {m}).
$$
\noindent Here, the internal state index \textit{B} is \textit{N} or $\Delta$. Using the
\textit{S}-matrix approach to the description of the $\gamma + \Delta \to N
+ \pi $ processes, transition operator $t_{\gamma \Delta \to N\pi}  $ was found
and it was presented as an expansion of four spin and three isospin
independent structures with the expansion coefficients that depend on the
coupling constants and magnetic moments. The single-particle transition
operator $t_{\gamma \Delta \to N\pi}$ is determined by the $\gamma + \Delta \to N
+ \pi $ process amplitude   \cite{10} that can be graphically represented as  
a sum of the diagrams shown in Fig. \ref{Fig:2}.

At the interaction of a photon with isobars $\Delta ^{ + +},\ \Delta ^{ +} $ and $\Delta ^{-}$
the amplitude corresponding to the diagram in Fig. \ref{Fig:2}\textit{a} dominates in the kinematic 
region of the $\Delta$(1232).
At the interaction of photon with neutral isobar $\Delta ^{0}$ this diagram does 
not contributes to the transition amplitude, but the amplitude corresponding to the contact diagram in Fig.
\ref{Fig:2}\textit{e} dominates.
As the single-particle transition operator $t_{\gamma N \to
N'\pi}  $, we will use the non-relativistic Blomqvist--Laget photoproduction
operator \cite{14}.

In accordance with (4),
\begin{equation}
\label{eq6}
\sum\limits_{f_{B}}  {\left| {T_{d}^{\Delta N}}  \right|^{2}} =
\sum\limits_{f_{B}}  {\left| {T_{d}^{\Delta} }  \right|^{2} + \left|
{T_{d}^{N}}  \right|^{2}} + \left| {T_{d}^{C}}  \right|^{2}.
\end{equation}

Considering the first term of this expression, we then in (\ref{eq5})  express the
wave function $\psi _{[ {\beta _{i} \beta _{j}}]}^{\Delta N}
( {x_{1} ,x_{2}})$ of the $\Delta$\textit{N} system through its
Fourier transform $\xi _{[ {\beta _{i} \beta _{j}} ]}^{\Delta N}
( {y_{1} ,y_{2}} )$, where $y \equiv {\bf p}, s, t$,
$$
\psi 
_{[ {\beta _{i} \beta _{j}}  ]}^{\Delta N}( {x_{1}
,x_{2}} ) =  
\frac{{1}}{{( {2\pi})^{3}}} \int {d(
{{\bf p}_{1},{\bf p}_{2}} )} \exp\left( {i ( {{\bf p}_{1} {\bf r}_{1} + {\bf p}_{2} {\bf r}_{2}}
)} \right)\, \xi _{[ {\beta _{i} \beta _{j}} ]}^{\Delta N}
( {y_{1} ,y_{2}})
$$
\noindent
and represent $\xi _{[ {\beta _{i} \beta _{j}} ]}^{\Delta N}
( {y_{1} ,y_{2}} )$ in the form of an expansion in states of the
$\Delta$\textit{N} system with total angular momentum \textit{J}, isospin
\textit{T}, the orbital angular momentum \textit{l}, total spin \textit{s}
and its projections \textit{M}$_{J}$, \textit{M}$_{T}$, \textit{m}$_{l}$ and
\textit{m}$_{s}$ as
\begin{equation}
\label{eq7}
\xi _{[ {\beta _{i} \beta _{j}}]}^{\Delta N} 
( {y_{1} y_{2}} ) = 
\sum\limits_{\beta _{S}}  {\Psi _{[ {\beta _{i} \beta
_{j}} ]  \beta _{S}} ^{\Delta N}( {\bf P},{\bf p})} \, \Omega
_{\beta _{S}} ( {s_{1} ,s_{2} ,t_{1} ,t_{2}}).
\end{equation}
\noindent Here, $\beta _{S}$ $ \equiv $ $ ( {\beta _{ex} \equiv J,M_{J} ,T,M_{T}}
),( {\beta _{in} \equiv l,s})$,
${\bf P} = {\bf p}_{1} + {\bf p}_{2} $ is
the momentum of the $\Delta$\textit{N} system,
${\bf p} = ( {\bf p}_{1} M_{N} - {\bf p}_{2} M_{\Delta} )
 /( M_{\Delta}  + M_{N})$
is the relative momentum, $M_{\Delta} $ and $M_{N} $ are isobar
and nucleon masses, respectively and $\Omega _{\beta _{S}}$  is the
spin--isospin wave function of the $\Delta$\textit{N} system.

As a result, after integration over internal, spin and isospin coordinates,
the squared modulus of the amplitude $T_{d}^{\Delta}  $ summed over the
states \textit{f}$_{B}$ of the residual nucleus \textit{B} and the spin
states \textit{f}$_{n}$ of the nucleon, can be written as
\begin{equation}
\label{eq8}
\sum\limits_{f_{B} ,f_{n}}  {\left| {T_{d}^{\Delta} }  \right|^{2}} =\, 
({2\pi})^{3} \mbox{Sp}( {t_{\gamma \Delta \to N\pi} \, \bm{\rho}
_{p} ^{\Delta}  ( {{\bf {p}}_{\Delta} } ) \, t_{\gamma
\Delta \to N\pi} ^{ \dagger} } ) 
\, \mbox{Sp}( {R_{m\tau _{\Delta} } ^{\Delta
} ( {{\bf {p}}_{\Delta} } )} ),
\end{equation}
\noindent
where
$$
\bm{\rho} _{p} ^{\Delta}  ( {\bf {p}} ) = \frac{{R_{m\tau
_{\Delta} } ^{\Delta}  ( {\bf {p}})}}{{\mbox{Sp}( {R_{m\tau _{\Delta}
}^{\Delta} ( {\bf {p}})} )}}
$$
\noindent is the polarization density matrix,
\begin{equation}
R_{m\tau _{\Delta} } ^{\Delta} ( {{\bf {p}}_{\Delta} }) = 
\sum\limits_{\beta _{ex} \beta _{in} \tilde {\beta} _{in}}  {\int {d{\bf {P}}}}
\, U_{\beta _{ex} \beta _{in} \tilde {\beta} _{in}}  ( {\hat {\bf {p}}}
) \, \rho _{m\tau _{\Delta}  ,\beta _{ex} \beta _{in} \tilde {\beta
}_{in}} ^{\Delta} ( {{\bf {P}},p}), \label{eq9}
\end{equation}
\[
U_{\beta _{ex} \beta _{in} \tilde {\beta} _{in}} ^{m\sigma _{\Delta}
m\tilde {\sigma} _{\Delta} }  ( {\hat {\bf {p}}}) =
\sum\limits_{m\sigma _{N}}  {S_{\beta _{ex} \beta _{in} \tilde {\beta} _{in}
}^{m\sigma _{\Delta}  m\tilde {\sigma} _{\Delta}  m\sigma _{N} m\sigma _{N}
} ( {\hat {\bf {p}}})} ,
\]
\[
\rho _{m\tau _{\Delta}  ,\beta _{ex} \beta _{in} \tilde {\beta} _{in}
}^{\Delta}  ( {{\bf {P}},{{p}}} ) = \sum\limits_{m\tau _{N}}  {\rho _{m\tau
_{\Delta}  m\tau _{N} ,\beta _{ex} \beta _{in} \tilde {\beta} _{in}}  (
{{\bf {P}},{{p}}})} ,
\]
\[
S_{\beta _{ex} \beta _{in} \tilde {\beta} _{in}} ^{m\sigma _{\Delta}
m\tilde {\sigma} _{\Delta}  m\sigma _{N} m\tilde {\sigma} _{N}}  (
{\hat {\bf {p}}} ) = \sum\limits_{m_{l} ,\tilde {m}_{l}}  {Y_{l,m_{l}}
( {\hat {\bf {p}}} )  Y_{\tilde {l},\tilde {m}_{l}} ^{\ast} (
{\hat {\bf {p}}})} \sum\limits_{m_{s} ,\tilde {m}_{s}}  {C_{l,m_{l}
;s,m_{s}} ^{J,M_{J}}    C_{{{3} \mathord{\left/ {\vphantom {{3} {2}}}
\right. \kern-\nulldelimiterspace} {2}},m\sigma _{\Delta}  ;{{1}
\mathord{\left/ {\vphantom {{1} {2}}} \right. \kern-\nulldelimiterspace}
{2}},m\sigma _{N}} ^{s,m_{s}}   }  C_{\tilde {l},\tilde {m}_{l} ;\tilde
{s},\tilde {m}_{s}} ^{J,M_{J}}    C_{{{3} \mathord{\left/ {\vphantom {{3}
{2}}} \right. \kern-\nulldelimiterspace} {2}},m\tilde {\sigma} _{\Delta}
;{{1} \mathord{\left/ {\vphantom {{1} {2}}} \right.
\kern-\nulldelimiterspace} {2}},m\tilde {\sigma} _{N}} ^{\tilde {s},\tilde
{m}_{s}}  ,
\]
\[
\rho _{m\tau _{\Delta}  m\tau _{N} ,\beta _{ex} \beta _{in} \tilde {\beta
}_{in}}  ( {{{\bf {P}},{{p}}}} ) = \sum\limits_{ij} {\psi _{[ {\beta _{i}
\beta _{j}} ]  \beta _{ex} \beta _{in}} ^{\Delta N}( {{{\bf {P}},{{p}}}}
) \, \psi _{[ {\beta _{i} \beta _{j}} ]  \beta _{ex}
\tilde {\beta} _{in}} ^{\Delta N\ast} ( {{{\bf {P}},{{p}}}})  \left|
{C_{{{3} \mathord{\left/ {\vphantom {{3} {2}}} \right.
\kern-\nulldelimiterspace} {2}},m\tau _{\Delta}  ;{{1} \mathord{\left/
{\vphantom {{1} {2}}} \right. \kern-\nulldelimiterspace} {2}},m\tau _{N}
}^{T,M_{T}} }  \right|^{2}} .
\]
\noindent
Here, $m\tau _{\Delta}  = m\tau _{\pi}  + m\tau _{n} $ is the third
projection of the isobar isospin,
${\bf {p}}_{\Delta}  = {\bf {p}}_{\pi}  + {\bf {p}}_{n} -
{\bf {p}}_{\gamma}  $ is the isobar momentum satisfying equality in (\ref{eq9})
\[
{\bf {p}} = {\bf {p}}_{\Delta}  - \frac{{M_{\Delta} } }{{M_{\Delta}  + M_{N}} }{\bf {P}}.
\]
In equation (\ref{eq7}), the wave function $\Psi _{[ {\beta _{i} \beta _{j}
} ]  \beta _{S}} ^{\Delta N} ( {{\bf {P}},{\bf {p}}} )$ is associated
with the function $\psi _{[ {\beta _{i} \beta _{j}}  ]  \beta
_{ex} \beta _{in}} ^{\Delta N} ( {{{\bf {P}},{p}}} )$ by the relation
\[
\Psi _{[ {\beta _{i} \beta _{j}} ]  \beta _{S}} ^{\Delta N}
( {{{\bf {P}},{\bf {p}}}}) = \psi _{[ {\beta _{i} \beta _{j}}]
\beta _{ex} \beta _{in}} ^{\Delta N} ( {{{\bf {P}},{p}}})  Y_{l,m_{l}}
( {\hat {\bf {p}}}).
\]
The trace of the matrix $R_{m\tau _{\Delta} } ^{\Delta}  $ is
$$
\mbox{Sp}  
( {R_{m\tau _{\Delta} } ^{\Delta}( {{\bf {p}}_{\Delta} })}
) = 
\frac{{1}}{{4\pi} }\sum\limits_{\beta _{S}}  {\int {d{\bf {P}}}}  \,
\rho _{m\tau _{\Delta}  ,\beta _{ex} \beta _{in} \beta _{in}} ^{\Delta}
( {{\bf {P}},{{p}}} ) = \rho _{m\tau _{\Delta} } ^{\Delta}  ( {{\bf {p}}_{\Delta
}} ).
$$
\noindent
Here, $\rho _{m\tau _{\Delta} } ^{\Delta}  ( {{\bf {p}}_{\Delta} } )$ is
the momentum distribution of the isobar in the charge state $m\tau _{\Delta
} + 0.5$.

Undertaking  a similar transformation for the second term of (\ref{eq6}), we obtain
\begin{equation}
\label{eq10}
\sum\limits_{f_{B} ,f_{n}}  {\left| {T_{d}^{N}}  \right|^{2}} = \, (
{2\pi})^{3}  
\mbox{Sp}( {t_{\gamma N \to N'\pi}  \, \bm{\rho} _{p} ^{N}
( {{\bf {p}}_{N}} )  t_{\gamma N \to N'\pi} ^{ +} }
) 
\times\,\mbox{Sp}( {R_{m\tau _{N}} ^{N} ( {{\bf {p}}_{N}} )} ),
\end{equation}
\noindent
where $m\tau _{N} = m\tau _{\pi}  + m\tau _{n} $ is the third isospin projection
of the nucleon of the $\Delta$\textit{N} system,
$$
\bm{\rho} _{p} ^{N} ( {\bf {p}} ) = \frac{{R_{m\tau _{N}} ^{N}
( {\bf {p}} )}}{{\mbox{Sp}( {R_{m\tau _{N}} ^{N} ( {\bf {p}} )}
)}},
$$	
\begin{equation}
\label{eq11}
\hspace*{-4mm} R_{m\tau _{N}} ^{N} ( {{\bf {p}}_{N}} ) = 
\hspace*{-4mm}\sum\limits_{\beta _{ex} \beta
_{in} \tilde {\beta} _{in}}  {\int {d{\bf {P}}}}  \, V_{\beta _{ex} \beta _{in}
\tilde {\beta} _{in}}  ( {\hat {\bf {p}}} )  \rho _{m\tau _{N} ,\beta
_{ex} \beta _{in} \tilde {\beta} _{in}} ^{N} ( {{\bf {P}},{{p}}}),	
\end{equation}
$$
V_{\beta _{ex} \beta _{in} \tilde {\beta} _{in}} ^{m\sigma _{N} m\tilde
{\sigma} _{N}}  ( {\hat {\bf {p}}} ) = \sum\limits_{m\sigma _{\Delta}
} {S_{\beta _{ex} \beta _{in} \tilde {\beta} _{in}} ^{m\sigma _{\Delta}
m\sigma _{\Delta}  m\sigma _{N} m\tilde {\sigma} _{N}}  ( {\hat {\bf {p}}}
)} ,
$$
$$
\rho _{m\tau _{N} ,\beta _{ex} \beta _{in} \tilde {\beta} _{in}} ^{N}
( {{{\bf {P}},{{p}}}}) = \sum\limits_{m\tau _{\Delta} }  {\rho _{m\tau
_{\Delta}  m\tau _{N} ,\beta _{ex} \beta _{in} \tilde {\beta} _{in}}  (
{{{\bf {P}},{{p}}}})} .
$$
The relative momentum \textbf{p} in (\ref{eq11}) is related to the nucleon momentum
${\bf {p}}_{N} = {\bf {p}}_{\pi}  + {\bf {p}}_{n} - {\bf {p}}_{\gamma}  $ 
of the $\Delta$\textit{N} system through the equation
\[
{\bf {p}} = - \, {\bf {p}}_{N} + \frac{{M_{N}} }{{M_{\Delta}  + M_{N}} }\,{\bf {P}}.
\]

Now consider the last term of  equation  (\ref{eq6}). For the \textit{p}-shell
nuclei having a large set of  nucleon states, the third component of the
density matrix (\ref{eq4}) can be written as
\begin{equation}
\label{eq12}
\rho ^{C}( {X_{1} ;\tilde {X}_{1}} ) 
=  \varphi _{N}({m_{1}} )
\left[ \frac{{N_C^{\Delta N}}}{{A}}
\sum\limits_{i=1}^A
\psi _{\beta _{i}} ( x_{1} )\, \psi _{\beta _{i}}^* (\tilde x_{1} )
\right] 
\varphi _{N}^{\ast} ( {\tilde {m}_{1}} ),
\end{equation}
\noindent
where
\[
N_C^{\Delta N} = N_N + N_{\Delta}\,\frac{A-2}{A}.
\]

Using (\ref{eq12}), we obtain
\begin{equation}
\label{eq13}
\sum\limits_{f_{B} ,f_{n}}  {\left| {T_{d}^{C}}  \right|^{2}}
= \, (
{2\pi} )^{3} \mbox{Sp}( {t_{\gamma N \to N'\pi} \, {} t_{\gamma N
\to N'\pi} ^{ \dagger} }) 
\frac{{1}}{{2\,\sigma _{N} + 1}} \, \rho
_{m\tau _{N}} ^{C} ( {{\bf {p}}_{N}}).
\end{equation}
\noindent
Here, $\sigma _{N} $ is the nucleon spin,
\begin{equation}
\label{eq14}
\rho _{m\tau _{N}} ^{C} ( {{\bf {p}}_{N}}) = N_{C}^{\Delta N}
\sum\limits_{i = 1}^{A} {\, \left| {\psi _{\beta _{i}}( {{\bf {p}}_{N}}
)} \right|^{2}{} } \,\delta _{m\tau _{N} ,m\tau _{i}}
\end{equation}
\noindent
is the momentum distribution of the nucleons with the third isospin
projection $m\tau _{N} $, constituting the nucleon core of the nucleus, and
$\psi _{\beta}  ( {\bf {p}})$ is the Fourier transform of the spatial
part of the wave function $\psi _{\beta}  ( {x} )$.

Equation (\ref{eq13}), for the squared modulus of amplitude $T_{d}^{C} $, differs
from similar expression obtained in the quasifree approximation, taking
into account the nucleon configurations of the nuclear wave function only,
by the presence of the factor $N_{C}^{\Delta N} $. For the $^{16}$O nucleus
factor $N_{C}^{\Delta N} $ is equal to 0.97.           

Now we consider the  quasifree pion photoproduction model, taking into
account the isobar configurations in the ground state of the nuclei.
In this approach, the independent particle model is used as a model of a
nucleus, and the isobars and nucleons in the nucleus are considered as independent
constituents. In this model the wave function of the isobar configurations
of a nucleus with closed shells can be represented as
\begin{equation}
\label{eq15}
\Psi _{\beta} ^{\Delta}  
( {X_{1} ,...,X_{A}})  = 
A_{\Delta  }   \sum\limits_{i} {\Psi _{\beta _{i}} ^{\Delta}  (
{X_{1}} ) \Psi _{( {\beta _{i}})^{ - 1}}^{N}(
{X_{2} ,X_{3} ,...,X_{A}})}
\end{equation}
\noindent
where $\Psi _{\beta} ^{\Delta} ( {X}) = \varphi _{\Delta}
( {m})\, \psi _{\beta} ^{\Delta} ( {x})$ is the
wave function of the virtual isobar in the nucleus in the $\beta$ state and  $\Psi
_{( {\beta})^{ - 1}}^{N} $ is the wave function of the nucleon
core including $A - 1$ nucleons, whose state can be described in terms of
the oscillator shell model.

If we use equation  (\ref{eq15}) for the wave function of the isobar
configurations, the squared modulus of the amplitude $T_{qf}  $ in the
quasifree approximation, summed over the residual nucleus states
\textit{f}$_{B}$ and the nucleon spin states \textit{f}$_{n}$, can be
written, as
\begin{equation}
\label{eq16}
\sum\limits_{f_{B} ,f_{n}}  {\left| {T_{qf} }  \right|^{2}} =
\sum\limits_{f_{B} ,f_{n}}  {\left| {T_{qf}^{\Delta} }  \right|^{2}} +
\left| {T_{qf}^{C}}  \right|^{2},
\end{equation}
\noindent
where
\begin{equation}
\label{eq17}
\sum\limits_{f_{B} ,f_{n}}  {\left| {T_{qf}^{\Delta} }  \right|^{2}} = \,
( {2\pi} )^{3}  
\mbox{Sp}( {t_{\gamma \Delta \to N\pi} \,
 t_{\gamma \Delta \to N\pi} ^{ \dagger} }) 
\frac{1}{2\,\sigma _{\Delta } + 1}\,
\rho _{m\tau _{\Delta} } ^{\Delta}
( {{\bf {p}}_{\Delta} } ),
\end{equation}
\[
\rho _{m\tau _{\Delta} } ^{\Delta}  ( {\bf {p}}) = \sum\limits_{i}
{\left| {\psi _{\beta _{i}} ^{\Delta}( {\bf {p}})} \right|^{2}{} }
\delta _{m\tau _{\Delta}  ,m\tau _{i}}  .
\]
\noindent
Here, $\sigma _{\Delta}  $ is the isobar spin and  $\psi _{\beta} ^{\Delta}
( {\bf {p}})$ is the Fourier transform of the spatial part of the
isobar wave function $\psi _{\beta} ^{\Delta}( {x})$.

The formula for the squared modulus of the amplitude $T_{qf}^{C} $ coincides
with (\ref{eq13}). The difference lies in the value of the coefficient that
determines the momentum distribution of the nucleons in the nucleus (\ref{eq14}).
The factor $N_{C}^{\Delta N} $ in (\ref{eq14}) should be replaced by
\[
N_C^{qf} = N_N + N_{\Delta}\,\frac{A-1}{A}.
\]

The coefficients $N_{C}^{qf} $ and $N_{C}^{\Delta N} $ differ by a value
$N_{\Delta} /  A$, which essentially does not exceed $\sim
$0.02 for the \textit{p}-shell nuclei \cite{6,11}.

\section{Results and discussion}

Depending on the charge state of the $\pi$\textit{N} pair, produced in the
$A( {\gamma ,\, \pi N})B$ reaction, a main contribution to
the cross section is given by different elements in equations (\ref{eq6}) and (\ref{eq16}).
In particular, the non-zero contribution to the production of the $\pi ^{ +
}p$ or the $\pi ^{ -} n$ pairs give only the first terms corresponding to
the interaction of a photon with the $\Delta^{++}$ or $\Delta^{-} $ isobars. In the case
of the $\pi ^{ +} n,\, \pi ^{0}p,\, \pi ^{0}n$ and $\pi ^{ -} p$ pair production
with an electric charge equal to 0 or +1, all terms in (\ref{eq6}) and (\ref{eq16})
contribute to the cross section. For example, at production of the $\pi ^{
-} p$ pair, the first terms in (\ref{eq6}) and (\ref{eq16}) correspond to the production of
the pion in the process $\gamma \Delta ^{0} \to \pi ^{ -} p$; the second
term in (\ref{eq6}), which  is absent in the quasifree pion production model,
corresponds to the production of the pion in the process $\gamma n \to \pi
^{ -} p$ at the interaction of the photon with neutron of the $\Delta ^{0}n$,
$\Delta ^{ +} n$ and $\Delta ^{ + +} n$ correlated systems. The last terms
in (\ref{eq6}) and (\ref{eq16}) describe the contribution to the cross section of the $\pi
^{ -} $ photoproduction on the neutrons of the nucleon core.

\begin{figure}
\begin{floatingfigure}{8cm}
\unitlength=1cm
\centering
\begin{picture}(8,8)
\put(0,2.7){\includegraphics[width=8cm,keepaspectratio]{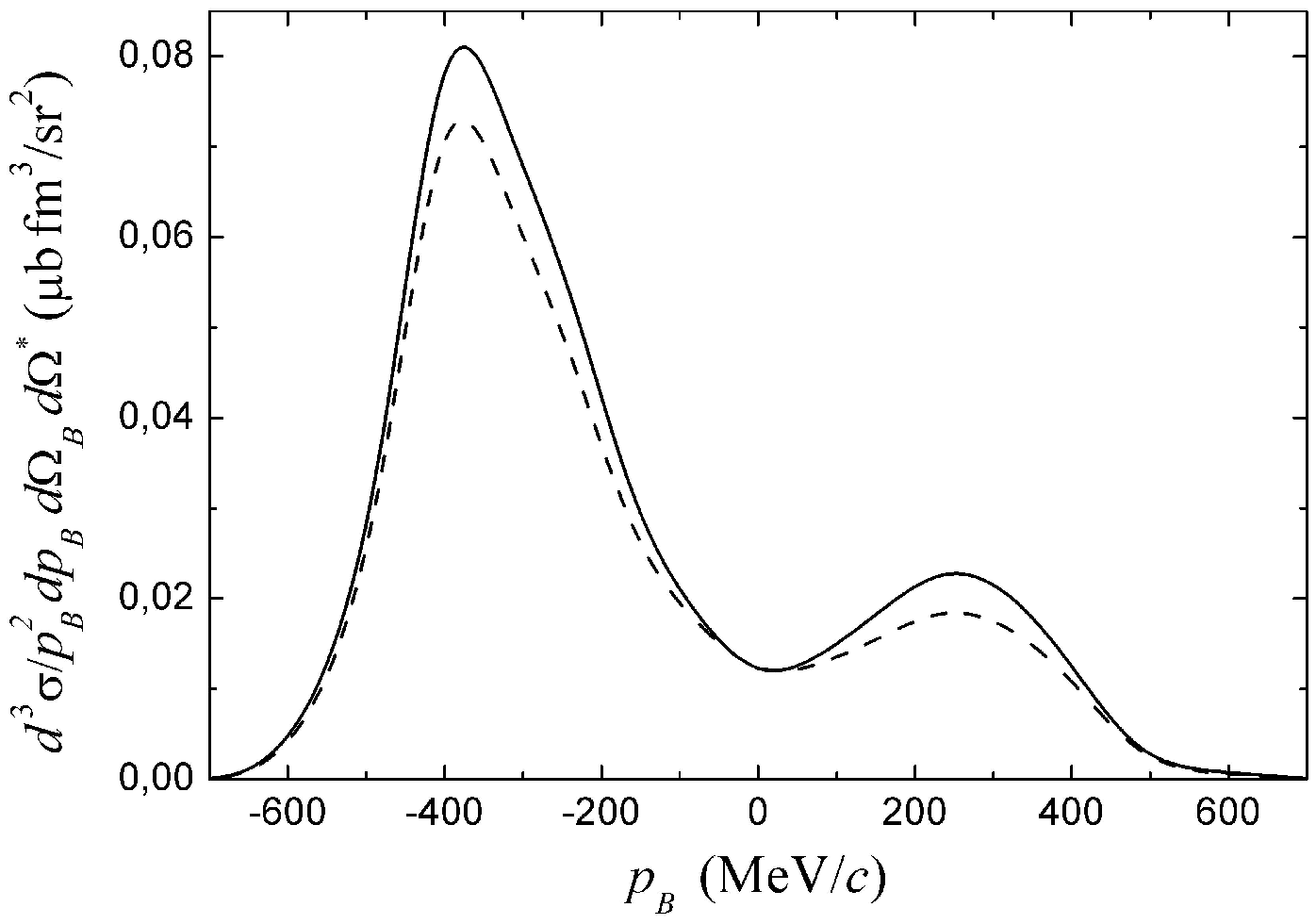}}
\end{picture}
\vspace*{-3.3cm}\caption{\label{Fig:3}\small Differential cross section of the reaction
$^{16}$O$( {\gamma ,\,\pi ^{ +} p}){}^{15}$C as a function of the momentum \textit{p}$_{B}$ of
the residual nucleus $^{15}$C at
$E_{\gamma}  = 450\,$MeV, $\theta _{\pi} ^{\ast}  = \varphi _{\pi}  = \theta _{B} = 90^\circ $.
The solid curve is the $\Delta$\textit{N} correlation model; the dashed curve is the quasifree
pion photoproduction model. The calculations were undertaken in plane-wave approximation.
} 
\vspace*{-0.8cm}
\begin{picture}(8,8)
\put(-0.3,1.2){\includegraphics[width=8cm,keepaspectratio]{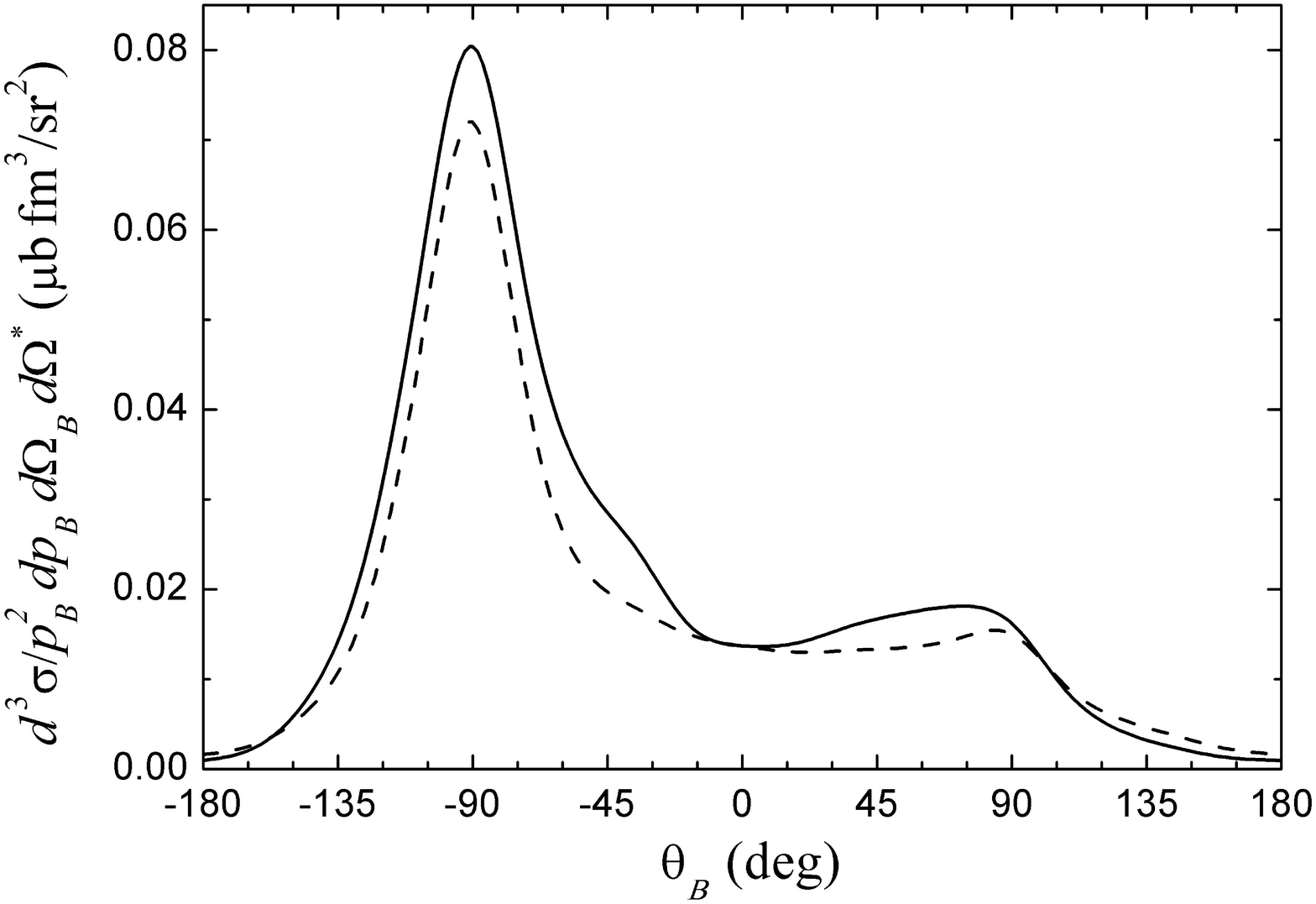}}
\end{picture}
\vspace*{-2cm}\caption{\label{Fig:4}\small Differential cross section of the reaction
$^{16}$O$( {\gamma ,\,\pi ^{ +} p})^{15}$C
as a function of the polar angle $\theta _{B} $ of the emission of the residual nucleus $^{15}$C at
$E_{\gamma}= 450\,$MeV, $\theta _{\pi} ^{\ast}  = \varphi _{\pi}  = 90^\circ $,
$p_{B} = 370\,\mbox{MeV}/c$. Designation of the curves is the same as in Fig. 3.
}
\end{floatingfigure}
\end{figure}
\vspace*{4mm}
Consider a prediction of the models for the two reactions $^{16}$O$( {\gamma
,\,\pi ^{ +} p}){}^{15}$C and $^{16}$O$( {\gamma ,\,\pi ^{ -} p}
){}^{15}$O.

The single mechanism of the direct production of the $\pi^{+}$\textit{p} pair
in the $( {\gamma ,\,\pi ^{ +} p})$ reaction is represented by the
diagram in Fig. 1\textit{a}. Comparing formulas (\ref{eq8}) and (\ref{eq17}) for the squared
modulus of the amplitude $T^{\Delta} $, we note that the expression \linebreak
Sp$( {t_{\gamma \Delta \to N\pi} \, \bm{\rho} _{p
}^{\Delta}  \,  t_{\gamma \Delta \to N\pi} ^{ \dagger} })$ in (\ref{eq8})
is the squared modulus of the matrix element of the transition $\gamma \Delta \to N\pi $, summed
over the spin states of the isobar and nucleon, which describes
the interaction of a photon with the polarized isobar. The polarization state of the isobar is
determined by the polarization density matrix $\bm{\rho} _{p}^{\Delta}  $.

The polarization density matrix $\bm{\rho} _{m\tau _{\Delta} } ^{\Delta}  $can be
expanded in the polarization operators \cite{15} and can be presented in its
simplest form as
$$
\bm{\rho} _{p} ^{\Delta}  = \frac{{1}}{{2\sigma _{\Delta}  +
1}}( {I + \Sigma}),
$$
\noindent
where $I$ is the unity matrix, having the dimension $( {2\sigma
_{\Delta}  + 1}) \times ( {2\sigma _{\Delta}  + 1} )$. If
the isobar is not polarized, the second term $\Sigma = 0$. Thus, ignoring
the effective polarization of the isobar in the initial state of the
process $\gamma \Delta   \to N\pi $, we obtain for the first term of (\ref{eq6}) a
natural transition from the model taking into account the $\Delta$\textit{N}
correlations, to an approach based on the independent particle model. The origin
of the effective polarization of the isobar in the $( {\gamma ,\,\pi N}
)$ process is related to the fact, that in the wave function expansion
(\ref{eq7}), the magnitude of the $\Delta$\textit{N} state contribution depends on the value
$m\sigma _{\Delta}  $. A degree of influence of the effective isobar
polarization on the cross section of the $^{16}$O$( {\gamma ,\,\pi ^{ +
}p}){}^{15}$C reaction can be estimated on the basis of the data
presented in Figs. 3, 4 and 5.

As we know, within the framework of the quasifree pion photoproduction,
polarization effects arise from the final state interaction \cite{16,17}. To
assess the effect of the polarization caused only by the $\Delta$\textit{N}
correlations, calculations were  performed in the plane-wave approximation.

Fig. 3 shows the dependence of the differential cross section of the
$^{16}$O$( {\gamma ,\pi ^{ +} p} ){}^{15}$C reaction plotted
against the momentum of the residual nucleus ${}^{15}$C. The calculations
were undertaken  in the following kinematic region:  photon energy of 450
MeV,  pion momentum in the c.m. of the pion-nucleon pair  perpendicular
to the photon momentum,  polar angle $\theta _{B} $ of the residual
nucleus emission and  azimuth angle $\varphi _{\pi}  $ of  pion
emission in the laboratory frame  equal to 90$^\circ $; it was assumed that
the geometry was coplanar and that the positive and negative values of the momentum on
the \textit{x}-axis corresponded to the azimuth angles $\varphi _{B} $ of
the emission of the residual nucleus ${}^{15}$C equal to $+90^\circ $ and
$-90^\circ $. The solid curve in Fig. 3 shows the reaction cross section
calculated using the $\Delta$\textit{N} correlation model, and the dashed curve is
the cross section calculated in the framework of the quasifree pion
photoproduction model. According to \cite{3,18}, it was assumed that the  
$\Delta$\textit{N} system produced upon the $NN \rightarrow \Delta N$ transition was
in a state, whose quantum numbers were \textit{J} = 0, \textit{T} = 1, and
\textit{l} = \textit{s} = 2. Significant asymmetry of the cross section with
respect to zero on the \textit{x}-axis is mainly due to an asymmetric
contribution of the diagram in Fig. 2\textit{a} to the transition amplitude
$\gamma \Delta ^{ + +}  \to p\pi ^{+} $.
\begin{figure}[t]
\begin{floatingfigure}{8cm}
\unitlength=1cm
\centering
\begin{picture}(8,8.5)
\put(0,2.5){\includegraphics[width=8cm,keepaspectratio]{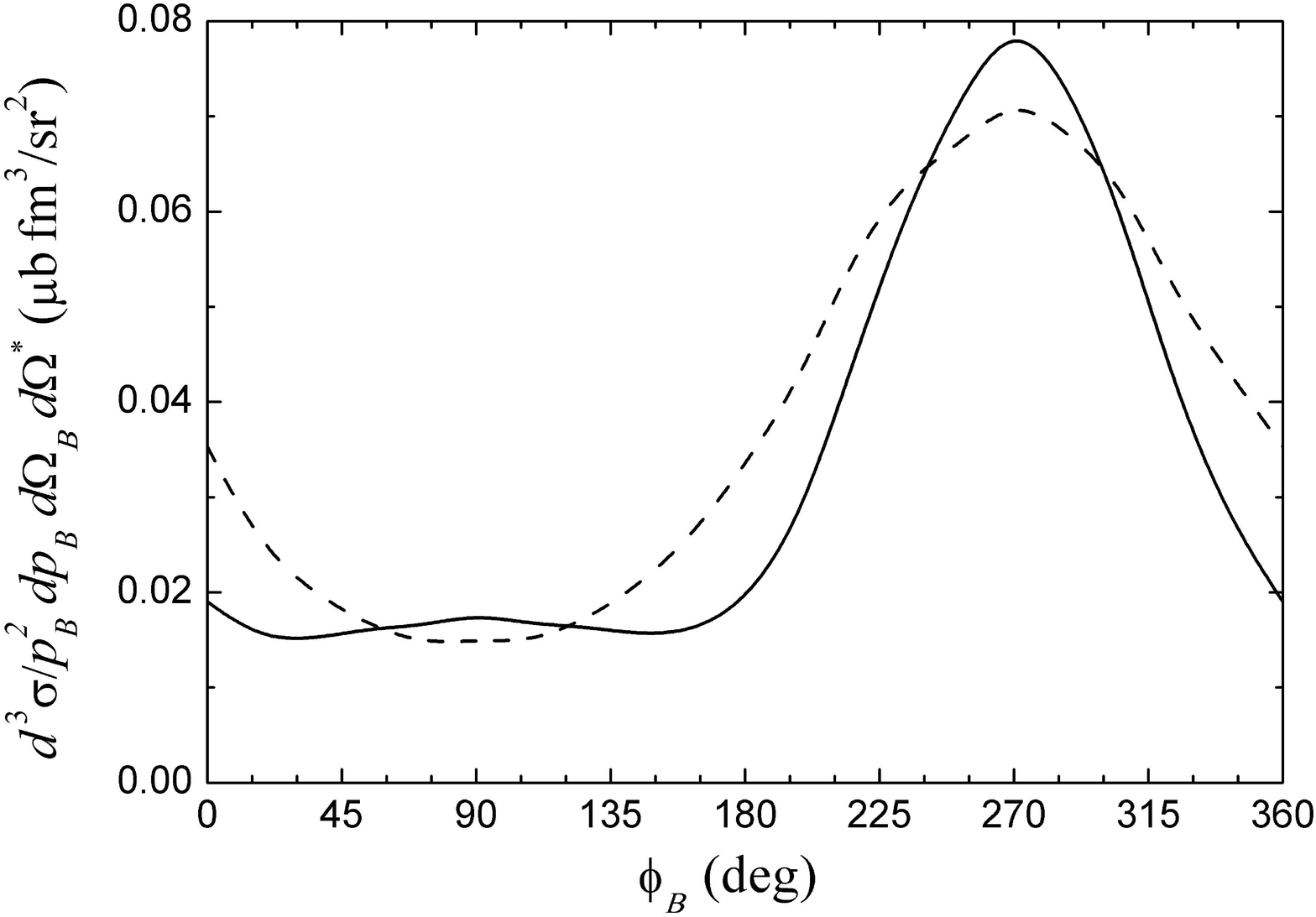}}
\end{picture}\\
\vspace*{-2.8cm}
\caption{\label{Fig:5}\small Differential cross section of the reaction
$^{16}$O$( {\gamma ,\,\pi ^{ +} p}){}^{15}$C as a function
of the azimuth angles $\varphi _{B} $ of the residual $^{15}$C nucleus
emission at $E_{\gamma}  = \mbox{MeV},\ \theta _{\pi} ^{\ast}  = \varphi _{\pi}  = 90^\circ,
\ p_{B} = 370\,\mbox{MeV}/c$. Designation of the curves is the same as in Fig. 3.
}
\end{floatingfigure}
\end{figure}

Figs. 4 and  5 show the dependences of the differential cross section of
the reaction $^{16}$O$( {\gamma ,\,\pi ^{ +} p} ){}^{15}$C plotted
against the polar $\theta _{B} $ and azimuth $\varphi _{B} $ angles of the
residual nucleus emission. The kinematic situation is different from the
previous case. In Fig. 4 the momentum of the residual nucleus is fixed and
it is equal to 370 MeV/\textit{c}, and in Fig. 5, additionally, the polar
angle $\theta _{B} $ is fixed and it equals 90$^\circ$. Positive and negative values
of the variable on the abscissa in Fig. 4 correspond to the azimuth angles
$\varphi _{B} $ of emission of the residual nucleus ${}^{15}$C as well as in
Fig. 3. In Fig. 5 the momenta of particles participating in the reaction are
coplanar when the azimuth angles $\varphi _{B} $ are 90$^\circ$ and 270$^\circ$.

As can be seen, outside the scope of the coplanarity of the particle
momenta, a difference in the differential cross sections of the
$^{16}$O$( {\gamma ,\,\pi ^{ +} p}){}^{15}$C reaction calculated
in the two models, reaches $\sim $ 80\%.

We now proceed to the analysis of the reaction $^{16}$O$( {\gamma ,\,\pi ^{ -} p}
){}^{15}$O.
\begin{figure}[h]
\vspace*{-5mm} 
\includegraphics[width=8cm,keepaspectratio]{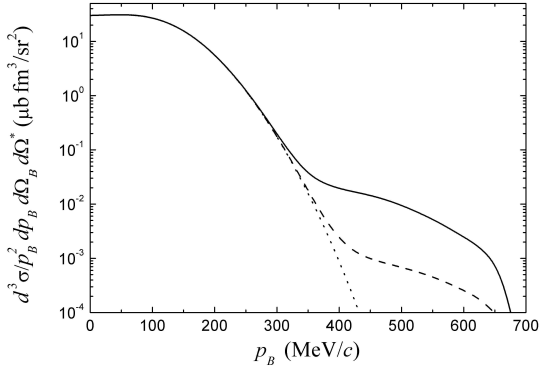}\\
\parbox[c]{8cm}{
\vspace*{1mm}\caption{\label{Fig:6}\small Differential cross section of the reaction
$^{16}$O$( {\gamma ,\,\pi ^{ -} p} )^{15}$O
as a function of the momentum $p_{B} $ of the residual nucleus
$^{15}$O at $E_{\gamma}  = 450\,\mbox{MeV},\
\theta _{\pi} ^{\ast}  = \varphi _{\pi}  = \theta _{B} = 90^\circ$,
$\varphi _{B} = - 90^\circ $. The solid curve is the $\Delta$\textit{N} correlation model, the dashed curve
is the quasifree pion photoproduction model and the dotted line is the contribution to the cross section
of the $T_{d}^{C} $ and $T_{qf}^{C} $ amplitudes corresponding to the interaction of the photons with the neutrons
of the nucleon core. The calculation was undertaken  in plane-wave approximation.
}
} 
\end{figure}
\begin{figure}[h]
\vspace*{-11.0cm}\hspace*{8.3cm} 
\includegraphics[width=8cm,keepaspectratio]{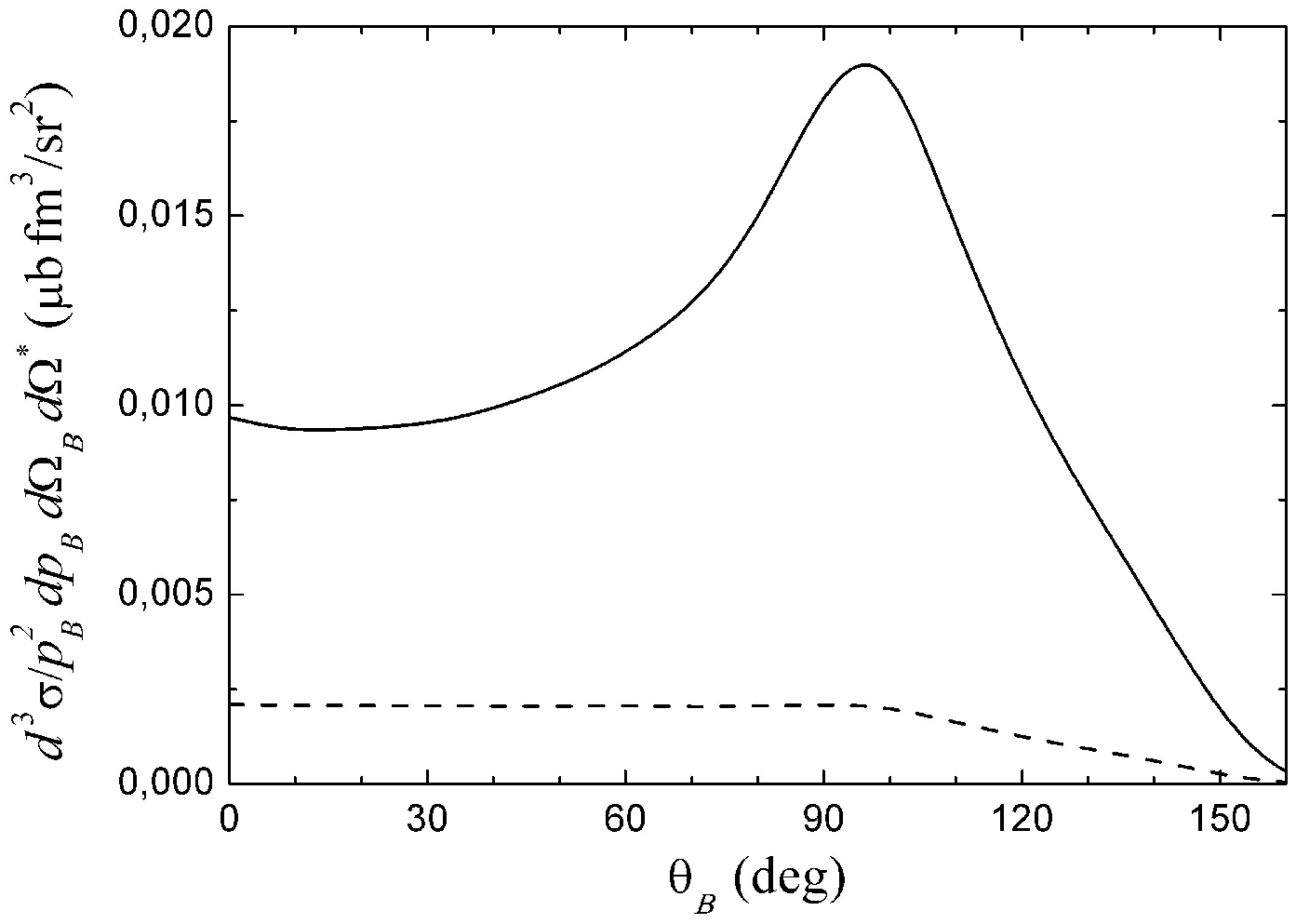}
\hspace*{8.3cm}
\parbox[t]{8cm}{
\vspace*{-3mm} \caption{\label{Fig:7}\small Differential cross section of the reaction
$^{16}$O$( {\gamma ,\,\pi ^{ -} p}){}^{15}$O as a function of the polar
angle $\theta _{B} $ of the residual nucleus $^{15}$O at $E_{\gamma}  = 450\,\mbox{MeV},
\ p_{B} = 400\,\mbox{MeV}/c, \ \theta _{\pi} ^{\ast}  = \varphi _{\pi}  = 90^\circ ,
\ \varphi _{B} = - 90^\circ .$ Designation of the curves is the same as in Fig. 6.
}
} 
\end{figure}

\vspace*{20mm}

In Fig. 6 we show the dependence of the differential cross section of the
$^{16}$O$( {\gamma ,\,\pi ^{ -} p}){}^{15}$O reaction plotted
against the momentum $p_{B} $ of the residual nucleus $^{15}$O at
$E_{\gamma}  = 450\,\mbox{MeV},\, \theta _{\pi} ^{\ast}  = \varphi _{\pi}  =
\theta _{B} = 90^\circ $, $\varphi _{B} = - 90^\circ $. 
The significant
difference in  process $( {\gamma ,\,\pi ^{ -} p})$ from
process $( {\gamma ,\,\pi ^{ +} p})$ is that the $\pi ^{ -} p$
pair production is possible at the interaction of a photon with the
nucleons. Therefore, the contribution to the reaction cross section is
possible from all terms in (\ref{eq6}) and (\ref{eq16}). 

\newpage
\begin{figure}[h]
\begin{floatingfigure}{8.0cm}
\unitlength=1cm
\centering
\begin{picture}(8,18.5)
\put(0.3,1.85){\includegraphics[width=7.6cm,keepaspectratio]{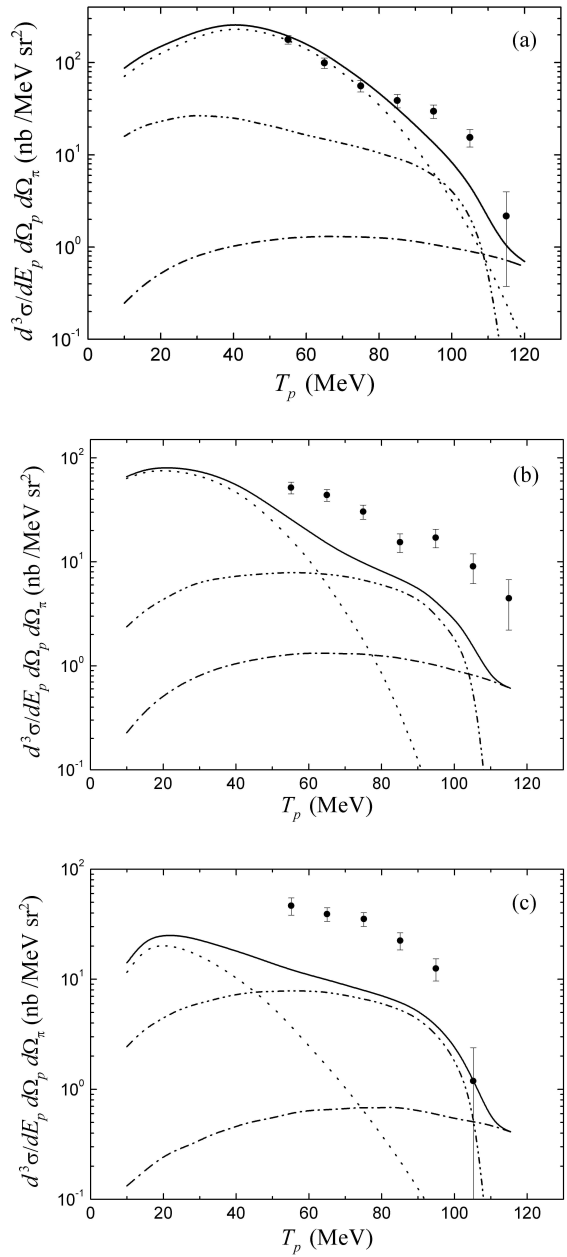}}
\end{picture}\\
\vspace*{-1.6cm}
\caption{\label{Fig:8}\small Differential cross section of the reaction $^{16}$O$( {\gamma ,\pi ^{ -} p})$
as a function of the kinetic energy of the proton \textit{T}$_{p}$. Data are taken from Ref. \cite{13}.
The dotted and dashed-dotted curves are the contributions to the cross section of the nucleon core
and the isobar configurations respectively, the dashed-dotted-dotted curve is the  sum of the
cross section of the $^{16}$O$( {\gamma ,\,\pi ^{ -} p})n{}^{14}$O and
$^{16}$O$( {\gamma ,\,\pi ^{ -} p})p{}^{14}$N reaction, the solid curve is the sum of
the cross section of the $\pi ^{ -} $ photoproduction with the one- and two-nucleons emissions.
The calculation was undertaken  in distorted wave approximation.
}
\end{floatingfigure}
\end{figure}

The dotted curve in Fig. 6 shows
the contribution to the cross section of the pion photoproduction on the
neutrons, corresponding amplitudes $T_{d}^{C} $ and $T_{qf}^{C} $, which,
under minor differences in magnitude, dominate at small momenta of the
residual nucleus. The solid and dashed curves show the differential cross
section calculated with the help of the $\Delta$\textit{N} correlation model and the
quasifree pion photoproduction model, respectively, taking into account the
isobar configurations in the ground state of a nucleus. As can be seen, at
momenta of the residual nucleus above $\sim $400 MeV/\textit{c}, the $\pi ^{
-} p$ pair production is almost entirely due to the isobar configurations and
 the differential cross sections, obtained in the framework of the two
models under consideration, differ in this kinematic region by almost one
order of magnitude.

Fig. 7. shows the differential cross section of the reaction $^{16}$O$(
{\gamma ,\,\pi ^{ -} p}){}^{15}$O plotted against polar angle $\theta
_{B} $ of the residual nucleus $^{15}$O at $E_{\gamma}  = 450\, \mbox{MeV},$
$p_{B} = 400\,\mbox{MeV}/c, \ \theta _{\pi} ^{\ast}  = \varphi _{\pi}  =
90^\circ , \ \varphi _{B} = - 90^\circ .$ As can be seen, the two models predict
the differential cross section, significantly differing both in an absolute
value  and in the shape of the angular dependence.

Currently, there are no experimental data for the reaction $A( {\gamma
,\, \pi N})B,$ that  would produce a  conclusion about the validity of
the considered reaction models. There are exclusive experimental cross
sections measured at a particular state of the residual nucleus \cite{19,20,21}.
However, these data were obtained in the region of small momenta of the
residual nucleus, where the contribution of the isobar configurations is
disparagingly small. There are experimental data measured in the region of
high-momentum transferred to the residual nucleus, but without restriction
on the missing energy \cite{13,22,23}. Therefore, these data include the final state
in which the residual nucleus is disintegrated. Such experimental data of
the $( {\gamma ,\,\pi ^{ +} p} )$ reaction measured at the Tomsk
synchrotron have recently been satisfactorily interpreted using the model of
reactions $( {\gamma ,\,\pi N})$ and $( {\gamma ,\,\pi NN}
)$ taking into account the $\Delta$\textit{N} correlations in the ground state
of the nuclei \cite{11}. Below we use this approach to  analyze the
$^{16}$O$( {\gamma ,\,\pi ^{ -} p})$ reaction data measured at the
BNL  LEGS \cite{13}.

Fig. 8 shows the differential cross section of the reaction $^{16}$O$(
{\gamma ,\,\pi ^{ -} p})$ plotted against kinetic energy of the proton
\textit{T}$_{p}$ at $E_{\gamma}  \simeq 300\,$MeV, (\textit{a}) $\theta
_{\pi}  = 44^\circ , \ \theta _{p} = 55^\circ;$ (\textit{b}) $\theta _{\pi
} = 36^\circ, \ \theta _{p} = 75^\circ ;$ (\textit{c}) $\theta _{\pi}  =
132^\circ , \ \theta _{p} = 75^\circ .$ We chose the cross section data from
the large amount of data obtained in the experiment in \cite{13}, in which the
average momentum $p_{B} $ transferred to the residual nuclear system is approximately
equal  to (\textit{a}) 200 MeV/\textit{c}, (\textit{b}) 300
MeV/\textit{c}, and (\textit{c}) 400 MeV/\textit{c}. In this range of the
momentum transfers, the differential cross section of the quasifree pion
photoproduction on the neutrons bound in a nucleus varies by more than an
order of magnitude.

The theoretical cross section shown in Fig. 8 was calculated using the
$\Delta$\textit{N} correlation model, which includes the direct and exchange
reaction mechanisms \cite{10,12}. The final state interaction was taken into
account in the optical model. The dotted curve shows the cross section of
the $\pi ^{ -} $ photoproduction in the reaction $^{16}$O$( {\gamma
,\,\pi ^{ -} p}){}^{15}$O on neutrons of the nucleon core (contribution
of the amplitude $T^{C}$). The dashed-dotted curve shows the contribution to the
cross section of the $^{16}$O$( {\gamma ,\,\pi ^{ -} p})^{15}$O
reaction of the isobar configurations in the ground state of the nucleus
$^{16}$O. The dashed-dotted-dotted curve is the sum of the cross sections of the
$^{16}$O$( {\gamma ,\,\pi ^{ -} p})n{}^{14}$O and $^{16}$O$(
{\gamma ,\,\pi ^{ -} p} )p{}^{15}$N reactions. In the kinematic region
considered above, the contribution of the isobar configurations in the
reaction $^{16}$O$( {\gamma ,\,\pi ^{ -} p})^{15}$O in the
framework of the quasifree pion photoproduction does not exceed 10$^{-1}$
nb/MeV sr$^{2}$. The solid curve shows the sum of the the $\pi ^{ -} $
photoproduction cross sections with the emission of one and two nucleons. As
can be seen, considering the isobar configurations, we satisfactorily
reproduced the form of the energy dependence of the reaction cross section.
However, disagreement of the absolute values of the experimental data and
theoretical cross sections increases with the growth in the momentum of the residual
nuclear system. At average momentum $p_{B} \approx 400$ MeV/\textit{c}, the
experimental differential cross section exceeds the
calculated cross section more than three-fold.

\section{Conclusions}

We have considered two models of the $A( {\gamma ,\, \pi N}
)B$ reaction that take into account the isobar configurations in the
ground state of an atomic nucleus: the $\Delta$\textit{N} correlation model and the
 quasifree pion photoproduction model. The main distinction between the two
models is the description of the state of an atomic
nucleus, which comprises the isobars. 
The general feature for these models is the approach,  
in  accordance with which isobars and nucleons are equal components of an 
atomic nucleus.  
Distinction between the models consists of the following: in the
$\Delta$\textit{N} correlation model, the dynamic relationship between the nucleon
and the isobar of the $\Delta$\textit{N} system, formed in the virtual transition $NN
\to \Delta N$, is taken into account. In the  quasifree pion
production model, the independent particle model is used -- nucleons and isobars in
a nucleus are considered independent. In the $\Delta$\textit{N} correlation model,
the photon interacts with baryons in three states: isobar, nucleon of the
$\Delta$\textit{N} system and nucleons of the nucleon core. In the quasifree pion
production model, there are only two such baryon states. This leads
to the main difference between the two models of the $A( {\gamma ,\,
\pi N})B$ reaction -- the additional amplitude $T_{d}^{N} $ (\ref{eq10}) in the
$\Delta$\textit{N} correlation model, which provides a significant contribution to
the cross section of the production of the pion--nucleon pair  with charge 0 and +1
at high momenta of the residual nucleus. Another difference between the
predictions of the two models is the presence of  
the polarization density matrix $\bm{\rho} _{p} ^{\Delta}$, 
in the expression for the squared modulus of the amplitude $T^{\Delta} $ (\ref{eq8}), 
which describes the interaction of a photon with the isobar 
within the $\Delta$\textit{N} correlation model. 
Effective polarization of the virtual isobar in the nucleus has a significant
impact on the value of the reaction cross section.

As is known, the quasifree pion photoproduction  model
satisfactorily describes the reaction $A( {\gamma ,\, \pi N} 
)B$ at sufficiently high momenta, transferred to the nucleon in the
process $\gamma N \to N'\pi $, and the small momenta of the residual nucleus,
where the contribution of the nucleon configurations dominates \cite{24,25,26}.
The independent particle model reproduces well the manifestations in the
reaction of the shell structure of the nucleus. However, a description of
the state of the nucleus, which includes the isobar configurations, within
the framework of the independent particle model, seems questionable. Because
of the short lifetime of the isobars, it is unlikely that, after its
appearance, the remaining $A - 1$ nucleons form a collective state with the
equilibrium momentum distribution, independent of the state of the isobar.
The $\Delta$\textit{N} correlation model eliminates this controversial hypothesis
by analyzing the state of the $\Delta$\textit{N} system, formed in the transition
$NN \to \Delta N$. In this model, the states of the nucleon and isobar of the
$\Delta$\textit{N} system are interdependent. Thus, the $\Delta$\textit{N} correlation
model is physically more justified.

Due to the current lack of experimental data for the exclusive $A(
{\gamma ,\, \pi N})B$ reactions at high momenta of the residual
nucleus, we cannot form an unambiguous conclusion about the validity of the
considered reaction models. However, a satisfactory description of the
$( {\gamma ,\pi ^{ +} p} )$ reaction data at the high momenta of
the residual nuclear system, with the help of the $\Delta$\textit{N} correlation
approach, is evidence in favor of the $\Delta$\textit{N} correlation model \cite{11}. 
As has been shown,  
the  $\Delta$\textit{N} correlation model  of 
the $( {\gamma ,\,\pi N})$ and $( {\gamma ,\,\pi NN} )$ reactions also
improves the description of the $^{16}$O$( {\gamma ,\,\pi ^{ -  
}p} )$ reaction data measured at BNL \cite{13}. 
The observed excess of the
experimental data of the $^{16}$O$( {\gamma ,\pi ^{ -} p} )$
reaction over the theoretical cross sections is connected, possibly, with a
contribution of reaction mechanisms to  the emission of two nucleons, which
are described by a model with two-body transition operators -- meson exchange
currents.

\section*{Acknowledgments}

This work was partly supported by the Competitiveness Enhancement Program of Tomsk Polytechnic University.

\end{document}